# Evidence for Strong Itinerant Spin Fluctuations
# in the Normal State of CeFeAsO$_{0.89}$F$_{0.11}$ Iron-Oxypnictides

F. Bondino[1], E. Magnano[1], M. Malvestuto[2], F. Parmigiani[1,3], M. A. McGuire[4], A. S. Sefat[4], B. C. Sales[4], R. Jin[4,5], D. Mandrus[4], E. W. Plummer[5], D. J. Singh[4] and N. Mannella[5,*]

[1]CNR-INFM, Laboratorio Nazionale TASC, S.S. 14, km 163.5, I-34012 Trieste, Italy.
[2]Sincrotrone Trieste S.C.p.A., Area Science Park, S.S. 14, km 163.5, I-34012 Trieste, Italy.
[3]Dipartimento di Fisica, Università degli Studi di Trieste, I- 34127, Trieste, Italy.
[4]Materials Science and Technology Division, Oak Ridge National Laboratory,
Oak Ridge, TN 37831, USA.
[5]Department of Physics and Astronomy, The University of Tennessee, Knoxville, TN 37996, USA
* To whom correspondence should be addressed. E-mail: nmannella@utk.edu

## ABSTRACT
The electronic structure in the normal state of CeFeAsO$_{0.89}$F$_{0.11}$ oxypnictide superconductors has been investigated with x-ray absorption and photoemission spectroscopy. All the data exhibit signatures of Fe $d$-electron itinerancy. Exchange multiplets appearing in the Fe 3s core level indicate the presence of itinerant spin fluctuations. These findings suggest that the underlying physics and the origin of superconductivity in these materials are likely to be quite different from those of the cuprate high-temperature superconductors. These materials provide opportunities for elucidating the role of magnetic fluctuations in high-temperature superconductivity.

The recent discovery of high-temperature superconductivity with critical temperature ($T_C$) now exceeding 55 K in iron-oxypnictides and related materials has generated enormous excitement in the community [1]. It is now possible to study high-temperature superconductivity and its relation to magnetism in a wide range of magnetic element-based materials. Cuprate high-temperature superconductors (HTSC) and Fe-based superconductors (FeSC) are the only families of compounds capable of exhibiting high-temperature superconductivity with $T_C$ exceeding 55 K. Despite an impressive body of recent theoretical and experimental work, it is still unknown whether the essential physics of FeSC can be considered similar to that of the cuprate HTSC. On the one hand, much like in the cuprates, superconductivity emerges in close proximity to a long-range-ordered antiferromagnetic ground state, leading credence to the central relevance of the interplay between superconductivity and spin-density-wave instability [2,3,4,5]. On the other hand, a recent study of SmFeAsO$_{0.85}$F$_{0.15}$ unveiled a superconducting gap whose nature is more consistent with a BCS prediction, not compatible with theoretical models specifically designed for the cuprate HTSC [6]. There is growing evidence of an interplay between magnetism and superconductivity in the FeSC, with a general phase diagram showing apparent competition between a magnetically ordered, spin density wave state and superconductivity. However, the nature of the interplay between the superconducting and magnetic states and in particular the interplay between fluctuations associated with the apparent proximity to magnetism and properties of the superconducting phase remains to be established.

A hallmark of cuprate HTSC which sets them apart from conventional BCS-like superconductors (SC) is the occurrence of unusual normal state properties which are outside the framework of Fermi-liquid theory. Most of these properties appear to be related to the effects of electron correlations induced through strong Coulomb interactions among electrons in the narrow

Cu $3d$ bands, in particular strong on-site Coulomb repulsions, related to the Mott-Hubbard insulating undoped phases. A description of the electronic structure in the normal state of FeSC is expected to elucidate the role of electron correlations, thus providing firm grounds of comparison for establishing commonalities and differences with the cuprate HTSC, and allowing the determination of the most appropriate starting point for describing the physics of this new class of materials.

Here, the electronic structure of the normal state of $CeFeAsO_{0.89}F_{0.11}$ is measured with photoemission spectroscopy (PES) and x-ray absorption spectroscopy (XAS). The Fe XAS and PES spectra do not display satellite features commonly found in the Cu spectra of cuprate HTSC, indicative of the absence of strong electron correlation and localization effects in the electronic structure. In sharp contrast to the cuprates HTSC, the Fe XAS and PES spectra exhibit spectral signatures which are typical of delocalized, itinerant electrons. The Fe $3s$ spectra show exchange multiplets due to the coupling of the final Fe $3s$ core hole state with the conduction band states, indicative of the presence of fluctuating spin moments on the Fe sites. These findings indicate that the FeSC must be considered a new class of materials, quite unlike the cuprate HTSC or conventional BCS SC.

Polycrystalline samples were grown with a standard solid-state synthesis [7]. The PES and XAS measurements were carried out on the BACH beamline at the Elettra Synchrotron Facility [8]. Several samples from different batches have been measured at room temperature in a pressure better than $4\times10^{-10}$ mbar after being fractured *in situ*. Quantitative PES analysis of core-level spectra confirmed the expected F content and showed minimal degrees of surface contamination [9].

The focus in this work is on Fe and Ce, the two elements possibly responsible for magnetism. The electronic structure of occupied and unoccupied Fe and Ce states was detected by means of PES and XAS, respectively (Fig. 1 and Fig. 2). The energy position and spectral line shape of the most intense multiplet lines ($\approx 882$ eV) in the Ce $M_{45}$ electron-yield XAS spectrum (Fig. 1(a)) provide a unique fingerprint for the presence of Ce predominantly in a +3 valence state (i.e. $4f^1$ ground state configuration) [10], fully consistent with magnetic susceptibility measurements in both $CeFeAsO_{0.89}F_{0.11}$ and CeFePO [11,12].

A direct comparison between the PES Fe $2p$ and Cu $2p$ core-level spectra reveals marked differences between the normal state electronic structure in FeSC and cuprate HTSC [Fig. 1 (b)]. While the Cu $2p$ spectrum in cuprate HTSC is characterized by satellite structures appearing at higher binding energies (BE) as a result of the localized character of the Cu $3d$ electrons, the Fe $2p$ core-level spectrum in $CeFeAsO_{0.89}F_{0.11}$ exhibits a main $2p_{3/2}$ peak at 706.5 eV which is markedly different from that of Fe ionic compounds [13], but more akin to that of Fe metal and intermetallic compounds [14]. Similar conclusions were reached from inspection of the Fe $L_{23}$ XAS spectrum [Fig. 1 (c)]. The absence of well-defined multiplet structures renders the line shape markedly different compared with those of $Fe^{2+}$ ionic oxides and much more similar to that of Fe metal [15], a feature indicative of delocalization of the $3d$ band states. Furthermore, the weak and broad shoulder at $\approx 709.5$ eV, which is absent in Fe metal but is present in the Fe XAS spectra of Fe–X (X is an *sp*-element) compounds with strong Fe $3d$-X n$p$ hybridization such as Fe silicides [16], is indicative of the covalent nature of the As and Fe conduction band states.

Fig. 2 (a) shows the valence band (VB) spectra at three different photon energies. Given that the samples are polycrystalline, these spectra provide a representation of the occupied density of states (DOS), modulated by atomic cross-section effects and instrumental resolution. The features at 0–0.6 eV and 4-6 eV are consistent with the DOS calculated for LaFeAsO [17,18]. The region between 0.6 and 4 eV is different from the calculated DOS since it is dominated by Ce 4f states, as shown below. The leading edge at the Fermi level ($E_F$) is exactly what is expected from the room temperature Fermi edge smeared by the instrumental resolution.

One of the most striking observations in this study is the nature of the Fe bands near $E_F$. There is a high DOS at $E_F$, in marked contrast to equivalent spectra from cuprate HTSC [cf. curve *d* in Fig. 2 (a)]. This is fully consistent with the difference between the Fe $2p$ and Cu $2p$ PES core-level spectra in Fig. 1(b): the high Fe DOS at $E_F$ is very effective in completely screening the Fe $2p$ core-hole excitation, leading to an absence in the core-level spectra of satellite structures.

Further insights into the character of the occupied states in CeFeAsO$_{0.89}$F$_{0.11}$ are provided by resonant PES (ResPES) measurements. In a ResPES measurement, the incident photon energy is tuned across the absorption edge of a deeper core level of an atom. The portion of the VB associated with just the resonating energy level(s) strongly localized on that atom is enhanced and thus extractable from the VB structure [19,20]. The clear enhancement of a structure ≈ 1-eV wide centered at a constant BE of ≈ 1.7 eV when the photon energy is tuned across the Ce 3$d$-4$f$ absorption edge allows unambiguous identification of the ≈ 1.7-eV portion of the VB with Ce 4$f$ states [Fig. 2 (b)], suggesting a high degree of localization of the Ce intermediate states.

A contrasting behavior is shown when VB spectra are measured with photon energies tuned across the Fe $L_{23}$ absorption edge [Fig. 2 (c)]. No resonance occurs close to $E_F$, an observation that may appear very puzzling at first, given that several band structure calculations place the Fe states at $E_F$. Nonetheless, this behavior is typical of metallic systems such as Ni, Cr, and Fe [19,20]. For photon energies lower than the absorption edge maximum (*EM*), the spectra show the enhancement of a structure located at constant BE of 4.2 eV identified with the so-called two-hole satellite, which occurs in the presence of a coherent resonance process with the resonance energy tracking the photon energy, known as radiationless resonant Auger scattering (RRAS) [19,20]. Starting with photon energies just below the *EM*, the resonance energy is converted into a constant kinetic energy feature now moving in a BE scale, indicating an inelastic process and loss of coherence, identified as a pure Auger behavior. The observation of a resonant Raman to normal Auger transition with characteristics very similar to those of metals provides further spectroscopic evidence for a delocalized character of the occupied Fe states in CeFeAsO$_{0.89}$F$_{0.11}$. The values of *EM* = 707.8 eV and *BE* = 706.5 eV are close to the ones measured in Fe [20]. Although this behavior is very similar to the one observed for metal Fe, we note two important differences; namely, (a) the *BE* of the two-hole satellite is 4.2 eV instead of 3.2 eV in Fe, and (b) the crossover from the RRAS regime to the Auger regime occurs in close proximity of the EM and not, as in Fe, ≈ 2 eV below it [20]. These differences suggest that, although the Fe states appear to be delocalized, they are not identical to the ones found in metallic Fe, which is expected because of the hybridization with the As states and the different Fe bonding topology.

The PES and XAS spectroscopic signatures of Fe $d$-electron itinerancy in CeFeAsO$_{0.89}$F$_{0.11}$ are indicative of marked differences with the electronic structure of cuprate HTSC. On the other hand, the occurrence of magnetism, whose nature in the normal state is particularly important because of its possible relevance to superconductivity, sets these materials apart from MgB$_2$ or BCS-like SC. A considerable number of FeSC with $T_C$ below 7 K, such as Fe silicates and rare-earth-filled skutterudites, show Pauli paramagnetic behavior in their normal states [21]. LaFePO exhibits paramagnetism in the normal state as well [21]. Band calculations suggest that the Fe 3$d$ electrons are in the low-spin configuration and that magnetism in these materials is itinerant in nature [17,18,21]. The magnetic susceptibility of CeFePO and CeFeAsO$_{0.89}$F$_{0.11}$ exhibits a Curie-Weiss behavior with an effective moment almost identical to the expected value 2.54 μ$_B$ for a free trivalent Ce ion [11,12]. Most experimental evidence points to the absence of a significant contribution of a paramagnetic Fe local moment in the normal state but at the same time indicates spin fluctuation effects.

In order to provide information on Fe magnetic states, we measured the Fe 3$s$ core-level spectra (Fig. 3). In transition metals, the 3$s$ core-level spectra can exhibit a multiplet splitting arising from the exchange coupling of the core 3$s$ electron with the net spin $S_v$ in the unfilled 3$d$ shell of the emitter atom [22]. For ionic compounds, the multiplet energy separation $\Delta E_{3s}$ permits estimating the net spin of the emitter atom, i.e., the local spin moment, via $\Delta E_{3s} = (2S_v + 1)J^{\text{eff}}_{3s\text{-}3d}$, where $J^{\text{eff}}_{3s\text{-}3d}$ denotes the effective exchange integral between the 3$s$ and the 3$d$ shells [22,23,24,25]. As the electronegativity of the ligand decreases, this exchange-splitting interpretation of the 3$s$ spectra is no longer complete since charge transfer final-state screening effects become important and can lead to additional spectral structures [24,25]. However, 3$s$ spectra in Fe-based materials are always found to have a considerable exchange-splitting component [24,25,26,27], which also clearly appears in CeFeAsF$_{0.11}$O$_{0.89}$ (Fig. 3).

A description of the 3$s$ spectra based on cluster models is very inadequate for systems of delocalized character such as CeFeAsO$_{0.89}$F$_{0.11}$. It has been shown that for itinerant systems such as Mn and Co, the values of the magnetic moment extracted from the $3s$ core-level spectra are remarkably close to the ones measured by Curie-Weiss type fits to magnetic susceptibility, ferromagnetic hysteresis loops and/or neutron scattering studies of ordered states. In these cases, the net spin $S_v$ was found by extrapolating the linear fit of the measured splitting $\Delta E_{3s}$ for ionic compounds versus $(2S_V + 1)$ [28,29,30]. Using the same approach for $CeFeAsO_{0.89}F_{0.11}$, the value of the measured splitting $\Delta E_{3s} \approx 3$ eV obtained with a two-component fit of the Fe $3s$ spectra gives a value of $2S_v + 1 \approx 2$ (Fig. 3). Interpreting this result as a measurement of the Fe spin moment, one obtains $\approx 1$ $\mu_B$. However, this interpretation is at odds with other measurements such as Mössbauer spectroscopy, NMR, and magnetic susceptibility, according to which in $CeFeAsO_{0.89}F_{0.11}$, Fe is nonmagnetic [11,12,31]. Although the extremely short time scale involved in the photoemission process ($10^{-17}$ - $10^{-16}$ s) can account for the disagreement with much slower ($10^{-7}$-s) Mössbauer and NMR measurements, it cannot explain the Ce-alone contribution to the Curie-Weiss susceptibility. In light of these considerations, the splitting of the Fe $3s$ level can be interpreted as evidence of Fe-based spin fluctuations. It is significant that similar occurrence is found in the intermetallic compound FeAl, whose Fe $3s$ core-level spectrum exhibits a two-peak structure although the compound is known to be nonmagnetic with fluctuating magnetism [14(b),32,33]. In particular, FeAl is a material that is thought to be near a magnetic quantum critical point where it is subject to strong itinerant spin fluctuations. Our interpretation is also consistent with the fact that the best fits to the Fe $3s$ spectra are always obtained when the curve fitting the peak at higher BE is mainly of Gaussian character, with a width much larger than that of the lower BE peak and that expected from experimental resolution. Indeed, a system with strong itinerant Fe-based spin fluctuations would locally mimic fluctuations in the moment on Fe sites, which should appear in an Fe $3s$ spectrum as sidebands at higher $BE$ with the envelope of the peaks being a Gaussian, reflecting the normal character of their distribution. This is different from spin-fluctuations associated with local moment magnetism as with the fast PES probe all Fe in the cast would exhibit the same large moment.

In summary, it appears that FeSC are quite unique—an itinerant Fe d-band character, a high DOS at $E_F$, no ferromagnetic ordering, and fluctuating magnetism all occur in the non-superconducting phase. The present results show by the example of $CeFeAsO_{0.89}F_{0.11}$ that strong itinerant spin-fluctuations are an important defining characteristic of the normal state of FeSC, and that high temperature superconductivity is possible without the signatures of strong local Mott-Hubbard type correlations that characterize cuprate HTSC.

Research sponsored by the Division of Materials Science and Engineering, Office of Basic Energy Sciences. Oak Ridge National Laboratory is managed by UT-Battelle, LLC, for the U.S. Department of Energy under Contract No. DE-AC05-00OR22725. Portions of this research performed by Eugene P. Wigner Fellows at ORNL. Discussions with I.I. Mazin and G. M. Stocks are gratefully acknowledged.

**Fig. 1.** Ce $M_{45}$ XAS, Fe $2p$ PES, and Fe $L_{23}$ XAS spectra measured in total electron yield (a) Ce $M_{45}$ XAS spectrum. (b) PES Cu $2p$ in cuprate HTSC (inset, from ref. [34]) and equivalent Fe $2p$ spectrum collected from $CeFeAsO_{0.89}F_{0.11}$ with 946-eV photon energy and 1-eV total resolution. Also visible is the F $1s$ spectrum. The absence of satellite structures, denoted by arrows, in the Fe PES spectra indicate that the core-hole excitation is completely screened by the Fe states at $E_F$. (c) Fe $L_{23}$ XAS spectrum.

**Fig. 2.** Valence band (VB) and ResPES spectra. (a) VB measured with photon energy of (A) 175.5, (B) 456.4, and (C) 703.2 eV and total instrumental resolution set to 0.08, 0.19, 0.34 eV, respectively. Also shown (D) is the VB spectrum of $YBa_2Cu_3O_{7-y}$ from ref. [35], showing the typical low DOS at $E_F$ in cuprates HTSC. (b)-(c) ResPES spectra across the Ce $M_5$ and Fe $L_3$ edges, respectively. Excitation energies are denoted in the insets. (b) The resonance at $\approx 1.7$ eV reveals unambiguously that the component at $\approx 1.7$ eV, absent in the DOS calculated for LaFeAsO, is due to Ce $4f$ states. (c) The 4.2-eV BE component in Fe ResPES displays a Raman to Auger behavior (see text).

**Fig. 3.** Fe 3*s* core level spectrum measured with a photon energy of 490.5 eV and a two-component fit providing a value of $\approx 2.9$ eV for the splitting. A Shirley-type background has been subtracted from the data. Inset: linear extrapolation of the value of the splitting versus $(2S_V +1)$ for several Fe compounds. The points denote the $\Delta E_{3s}$ values for $FeF_3$ (marker: ▲) [36], $FeF_2$ (marker: ●) [37], FeO (marker: ✳) [25], Fe metal (marker: ✖) [26] and $CeFeAsO_{0.89}F_{0.11}$ (marker: ■) (this work). The continuous line denotes the linear fit resulting in a relation of the multiplet splitting vs. (2S+1) as $\Delta E_{3s} = 0.94 + 1.01 \times (2S+1)$.

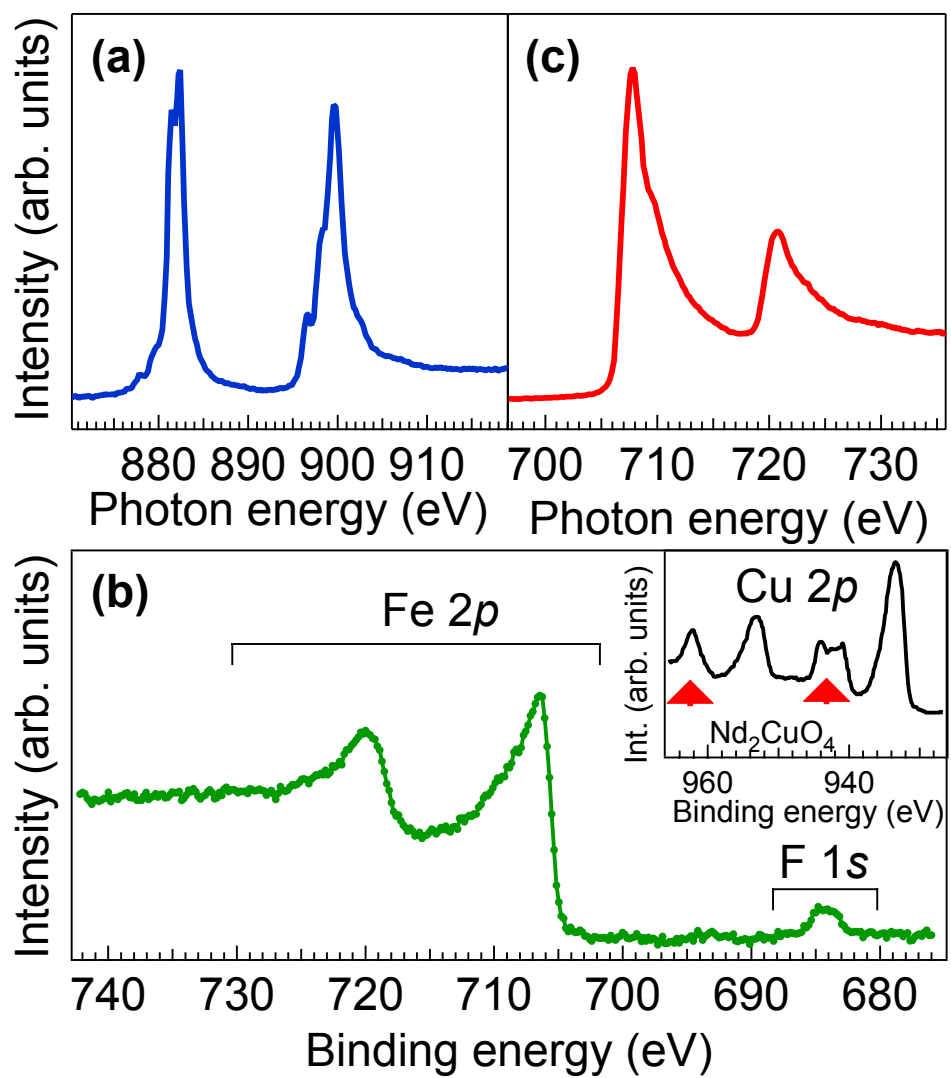

Figure 1

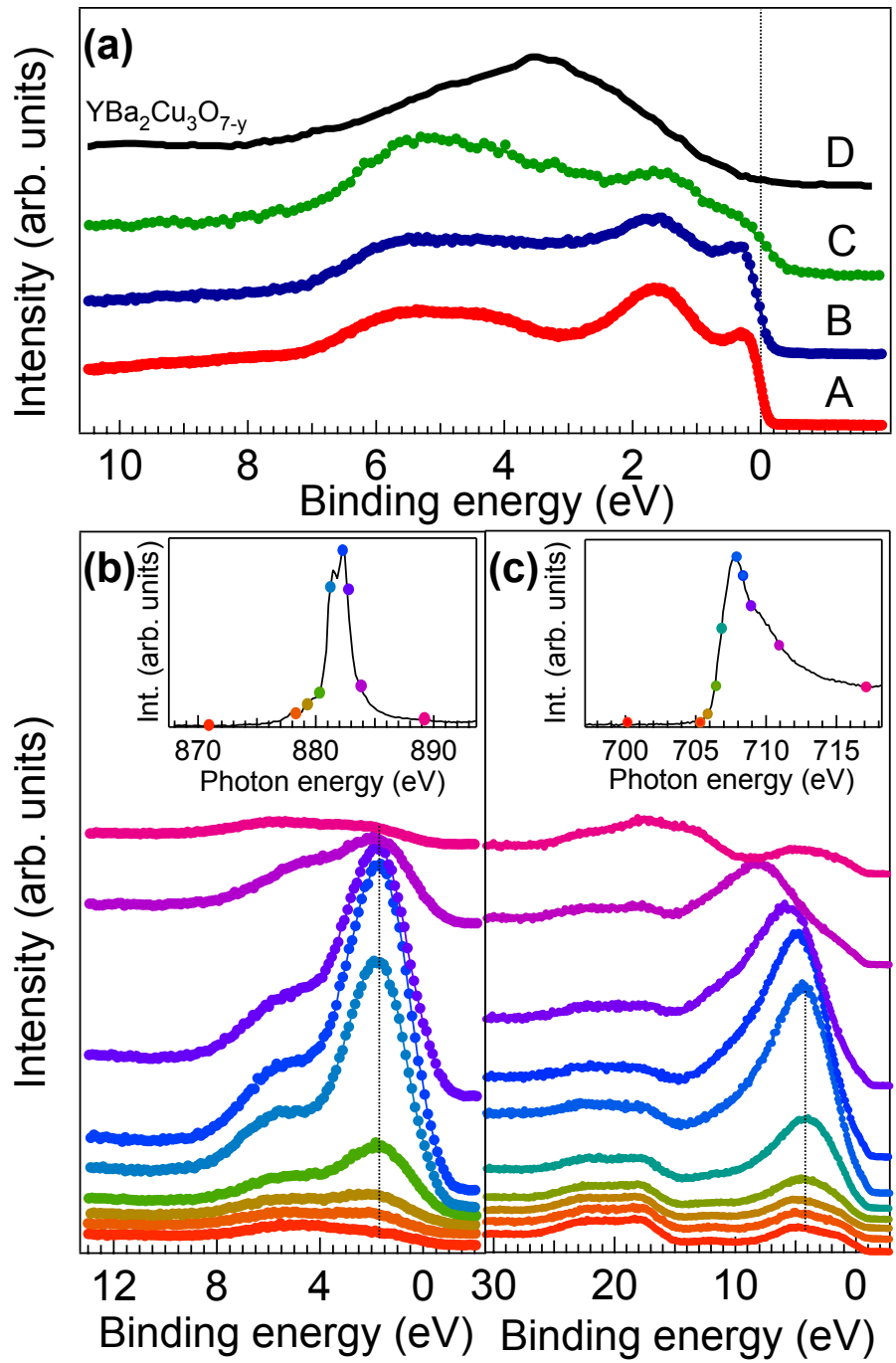

Figure 2

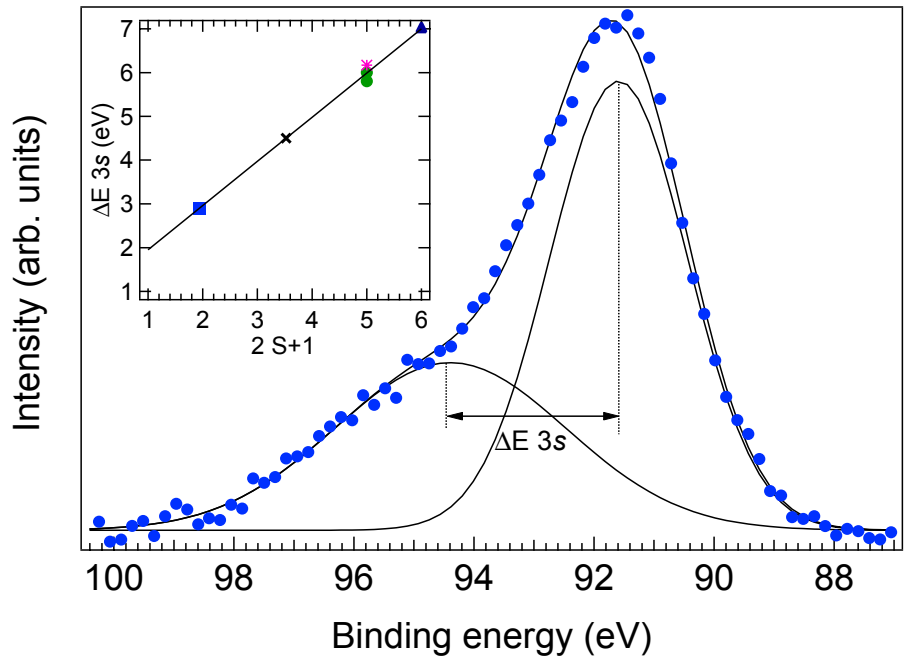

Figure 3